\documentclass[apj]{emulateapj}
\usepackage{booktabs} 
\usepackage{amsmath}
\usepackage{xcolor}
\shorttitle{VLBI astrometry of PSR B1913+16}

\newcommand{\psrpi}{\ensuremath{\mathrm{PSR}\Large\pi}}

\newcommand{\uas}{\ensuremath{\mu\mathrm{as}}}
\newcommand{\degrees}{\ensuremath{^\circ}}

\begin{document}
\title{A VLBI distance and transverse velocity for PSR B1913+16}

\author{
A. T. Deller\altaffilmark{1},
J. M. Weisberg\altaffilmark{2},
D. J. Nice\altaffilmark{3}, and
S. Chatterjee\altaffilmark{4}}
\altaffiltext{1}{Centre for Astrophysics and Supercomputing, Swinburne University of Technology, Mail Number H11, PO Box 218, Hawthorn, VIC 3122 Australia}
\altaffiltext{2}{Department of Physics and Astronomy, Carleton College, Northfield, MN 55057}
\altaffiltext{3}{Department of Physics, Lafayette College, Easton, PA 18042, USA}
\altaffiltext{4}{Cornell Center for Astrophysics and Planetary Science, Cornell University, Ithaca NY 14853, USA}

\begin{abstract}
Using the Very Long Baseline Array, we have made astrometric observations of the binary 
pulsar B1913+16 spanning 
an 18 month period in 2014 -- 2015.  From these observations we make the first determination of 
the annual geometric parallax of B1913+16,  measuring $\pi= 0.24^{+0.06}_{-0.08}$ milliarcseconds (68\% 
confidence interval).  The inferred parallax probability distribution differs significantly from a Gaussian.
Using our parallax measurement and prior information on the spatial and luminosity distributions of the millisecond pulsar population,
we infer a distance of $d=4.1^{+2.0}_{-0.7}$ kpc, which is significantly 
closer  than 
the $9.8 \pm 3.1$ kpc suggested by the pulsar's dispersion measure and analyses of the 
ionized interstellar medium.  
While the relatively low significance 
of the parallax detection 
($\sim$3$\sigma$) currently precludes an improved test of general relativity using the orbital decay of 
PSR B1913+16, ongoing observations with improved control of systematic astrometric errors 
could reach the 10\%  distance uncertainty required for this goal.  The proper motion measured 
by our VLBI astrometry differs substantially from that obtained by pulsar timing, a discrepancy 
that has also been found between the proper motion measurements made by interferometers 
and pulsar timing for some other pulsars and which we speculate is the result of timing noise or 
dispersion measure variations in the timing data set. Our parallax and proper motion measurements yield a transverse velocity
of $15^{+8}_{-4}$ km s$^{-1}$ in the solar reference frame.  Analysis incorporating galactic rotation and solar motion 
finds the space velocity of the pulsar relative to its standard of rest has a component 22$^{+7}_{-3}$ km s$^{-1}$
perpendicular to the galactic plane and components of order 100 km s$^{-1}$ parallel to the galactic plane.

\end{abstract}


\section{Introduction}
 
The double-neutron-star system containing PSR B1913+16 was the first binary pulsar system discovered
\citep{hulse75a} and has proved to be 
an outstanding laboratory for the study of
relativistic gravitation, owing to the extreme physical conditions characterizing it.
The observation of orbital decay at the rate expected from general relativity (GR) provided the first 
evidence for 
gravitational radiation emission \citep{taylor82a}, and improved observational data has led to 
ever-tighter constraints on the agreement with GR predictions \citep{weisberg16a}.

As the measurement precision has improved,  it has become  important to account for  
non-GR contributions to the observed orbital period derivative $\dot{P}_{\rm b}^{\rm obs}$.  
\citet{damour91a} pointed out, for example, that $\dot{P}_{\rm b}^{\rm obs}$
is affected not only by gravitational radiation emission but 
also by acceleration in the galactic gravitational 
potential and by 
the apparent acceleration imposed by the pulsar's transverse motion \citep{shklovskii70a}. 
Until now, our knowledge of these contributions, and hence the accuracy of 
the $\dot{P}_{\rm b}$ test of gravitational radiation and general relativity, has been limited primarily 
by the uncertainty in
the pulsar distance; and to a lesser degree, by the proper motion uncertainty. The pulsar distance has 
previously been  best constrained by  careful modeling of the galactic electron density along the line of sight toward it, and then
determining a distance based on the pulsar's dispersion measure  \citep{weisberg08a}.  
 The {\it{timing}} signature of annual parallax is too small
to be measured with the currently available precision, while the timing signature of proper motion 
may be distorted by any unmodeled influences on measured pulse arrival times, such as rotation 
irregularities (``timing noise'') or variations in interstellar dispersion not accounted for in the timing analysis.  
Accordingly, these quantities must be 
obtained via another procedure.  
Therefore, in an effort to improve the $\dot{P}_{\rm b}$ test, we embarked on an 
Very Long Baseline Interferometry (VLBI) astrometric program to measure the parallax and proper motion of this 
pulsar.  

Differential VLBI astrometry (measuring the position of a target source relative to the reference position of a nearby
calibrator) has been employed to measure the proper motion and parallax of dozens of radio pulsars
\citep[e.g.][]{deller09b,chatterjee09a}.  When the calibrator--target angular separation is small 
($\ll 1$\degrees), the
calibrator and target source can be observed simultaneously in a single pointing at the low frequencies where pulsars 
are brightest, in which case this differential offset can typically be measured with sub-milliarcsecond precision.  
By observing a number of times over a period of 1--2 years, effects such as proper motion and annual geometric 
parallax that change this offset can be measured to very high accuracy.  

In this paper, we describe our VLBI measurements (\S\ref{sec:vlbi}) of PSR B1913+16 and use them to obtain independent 
measurements of 
the pulsar's distance and transverse velocity (\S\ref{sec:astrometricresults}).   In \S\ref{sec:discussion}, we compare these 
results to other constraints on these parameters, and obtain a best estimate for the kinematic contributions to 
$\dot{P}_{\rm b}^{\rm obs}$ which we use to revise the gravitational radiation test with PSR B1913+16.  Our conclusions
are in \S\ref{sec:conclusions}.

\section{VLBI observations and data processing}
\label{sec:vlbi}
\subsection{Observations}
\label{sec:vlbiobs}
All observations of PSR B1913+16 were performed with the Very Long Baseline Array (VLBA).
Before commencing astrometry, we first sought a suitable ``in-beam" calibrator source by 
making snapshot VLBA images of all sources brighter than 3 mJy within 30\arcmin\ of PSR B1913+16,
to identify sources that were sufficiently bright and compact on VLBI scales.
This observation took place in 2011 May as a part of the \psrpi\ campaign 
\citep{deller11b,deller16a} and identified the 6 mJy source 
J191621.8+161853, which is separated by 18\arcmin\ from PSR B1913+16, to be sufficiently compact.  

Between 2014 May and 2015 November, eight astrometric observations were made with the VLBA (project code BD178).  
Two observations were made every six months, close to the parallax extrema in  April and  October. Each observation was 
4 hours in duration, during which time phase referencing was performed between the target (a pointing center midway 
between PSR B1913+16 and J191621.8+161853) and a primary phase calibrator at an angular separation of 
1.1\degrees\ (J1911+1611).  The phase reference cycle time was 4 minutes, with 3 minutes spent on the target and 
1 minute on the phase reference source.  To calibrate the instrumental bandpass, we observed the bright ``fringe finder" 
source J1925+2106 once per observation.  The maximum recording rate of 2 Gbps was used, sampling a total bandwidth 
of 256 MHz in two circular polarizations.  Eight 32 MHz subbands distributed between  1392 to 1744 MHz were used, 
avoiding regions of the radio spectrum most contaminated by radio frequency interference.  Correlation was performed using the DiFX correlator 
in Socorro \citep{deller11a}, with two correlation passes used to provide separate datasets centered on 
J191621.8+161853 and on PSR B1913+16.  On-pulse gating 
was used to improve sensitivity on PSR B1913+16, with the pulsar duty cycle of 12\% 
leading to a factor of $1 / \sqrt{0.12} \simeq 3$ improvement in signal--to--noise.

\subsection{Data reduction}
\label{sec:vlbiproc}
We used the same astrometric data reduction pipeline described in \citet{deller13a} and refined in 
\citet{deller16a}.  Briefly, this ParselTongue \citep{kettenis06a} script applies \textit{a priori} flags and user-supplied flags, 
applies \textit{a priori} amplitude calibration, corrects for ionospheric propagation delays, 
and then derives and applies successive calibration corrections using the ``fringe finder" source (time-independent 
delay and bandpass), the primary phase reference source (time dependent delay, amplitude, and phase), and the 
nearer in-beam calibrator (phase only).  For the ionospheric correction, the global model  
(grid resolution 2.5\degrees\ in latitude, 5\degrees\ in longitude, and 2h in time) 
produced by the International GNSS Service\footnote{available  
from \url{ftp://cddis.gsfc.nasa.gov/gps/products/ionex/}} was used.
The phase solutions on the in-beam calibrator were derived with a solution 
interval of 2.2 minutes, short enough to allow 2 solutions per scan on the target, but long enough to minimise the 
number of failed solutions.  After all calibration was applied, the target data 
were split and exported for imaging in difmap \citep{shepherd97a}.

The imaging and position extraction was identical to that performed in \citet{deller16a}, fitting a single point source model to the visibility data and making a single Stokes $I$ image per epoch.  An astrometric position and uncertainty was obtained from each resultant image by using the task JMFIT in AIPS \citep{greisen03a} to fit an elliptical Gaussian to the source.  With a typical synthesized beam size of $5\times10$ milliarcseconds (mas) at position angle $\sim$0\degrees\ and a typical detection significance of $\sim$40$\sigma$, the formal fit uncertainties at each epoch were $\sigma\mathrm{RA}_f \sim 0.05$ mas, $\sigma\mathrm{Decl}_f \sim 0.11$ mas in right ascension and declination respectively. We also split each observation into 30 minute chunks and extracted position fits for each chunk separately, to assist with the estimation of systematic position shifts due to the ionosphere as described below.

\section{Astrometric results}
\label{sec:astrometricresults}
\begin{figure}
\begin{center}
\begin{tabular}{c}
\includegraphics[width=0.49\textwidth]{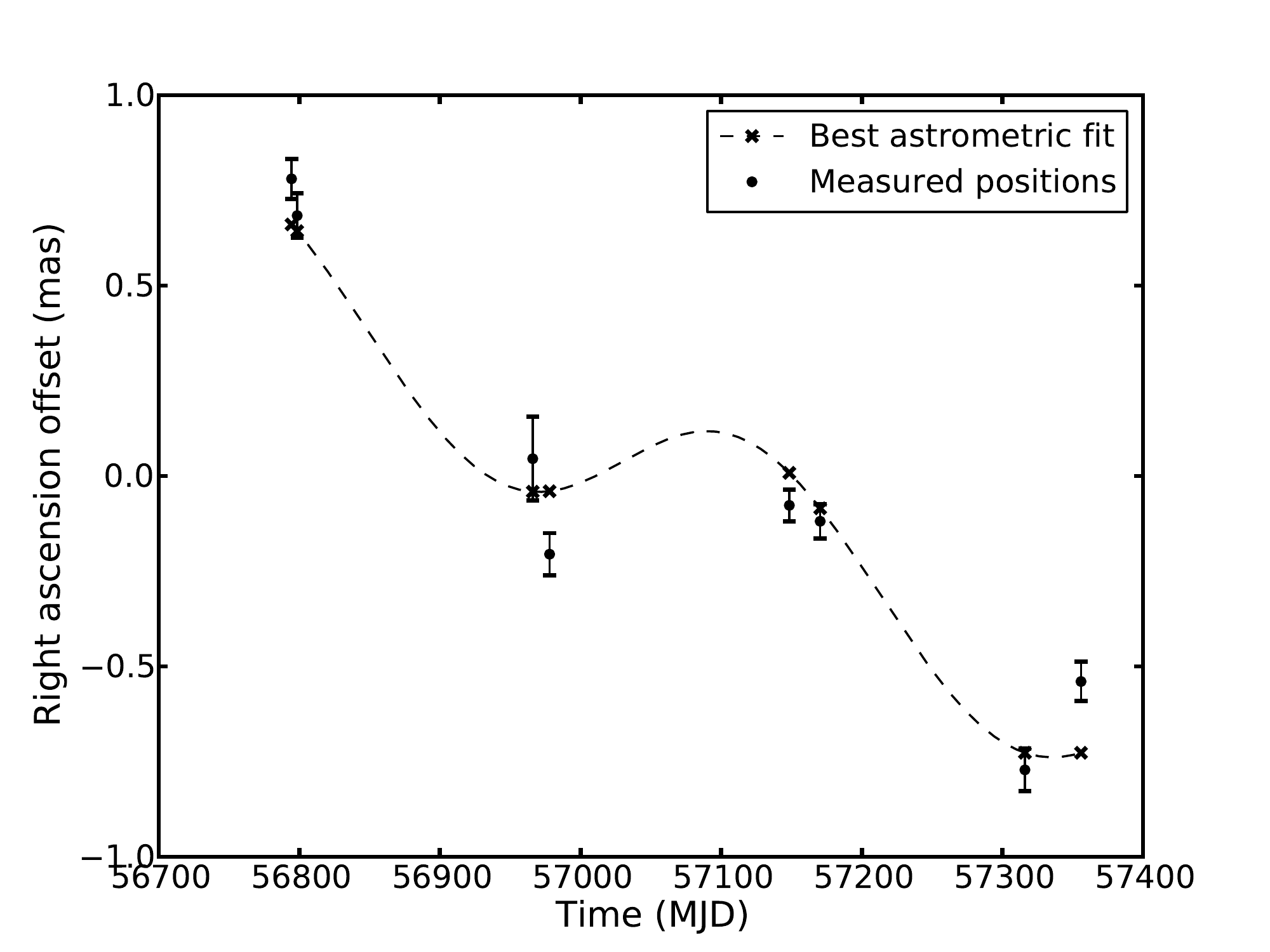} \\
\includegraphics[width=0.49\textwidth]{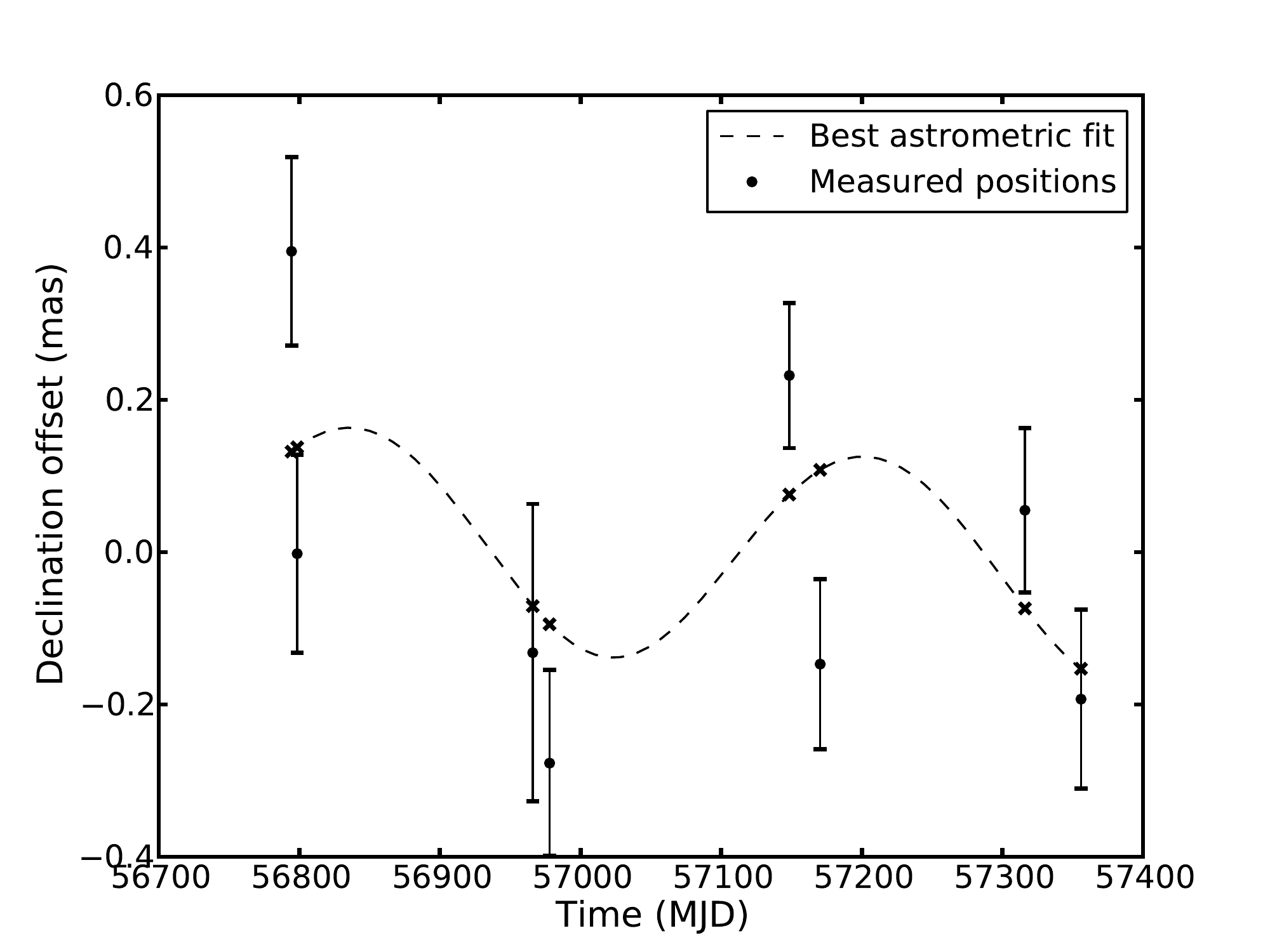} 
\end{tabular}
\end{center}
\caption{\label{fig:simplefit}The least-squares astrometric fit for PSR B1913+16, where the top panel and bottom panels show offset from the reference position (at MJD 57050) in right ascension and declination respectively. Error bars show formal position fit uncertainties only, which clearly underestimate the total astrometric error.  The reduced $\chi^2$ of the fit is 4.6.}
\end{figure}

The uncertainties on the fitted positions $\sigma\mathrm{RA}_f$, $\sigma\mathrm{Decl}_f$ that we obtain are purely determined by the interferometer resolution and the signal--to--noise ratio of the image, and do not account for potential systematic offsets introduced by the differential ionosphere between the in-beam calibrator and the target pulsar, or by structure evolution in the calibrator source.  Such systematic errors $\sigma\mathrm{RA}_{\mathrm{sys}}$, $\sigma\mathrm{Decl}_{\mathrm{sys}}$ are expected to be comparable to or greater than $\sigma\mathrm{RA}_f$ and $\sigma\mathrm{Decl}_f$ based on the nominal accuracy of the global ionospheric models used, and on previous experience of pulsar astrometry at these frequencies \citep[e.g.,][]{chatterjee09a}.  This is confirmed by making a simple least-squares fit of reference position, proper motion, and parallax to our measured positions using only the formal position fit uncertainties $\sigma\mathrm{RA}_f$, $\sigma\mathrm{Decl}_f$.  The rms of the residual position offsets is 0.11 mas in right ascension and 0.17 mas in declination, leading to a $\chi^2$ of 51 for 11 degrees of freedom.  This fit is shown in Figure~\ref{fig:simplefit}.

If this simple fit were taken at face value, we would obtain a parallax for B1913+16 of $0.22 \pm0.02$ mas, where the uncertainty is obviously under-estimated.  In order to obtain realistic uncertainties on the astrometric parameters, we need to account for the systematic error contributions $\sigma\mathrm{RA}_{\mathrm{sys}}$ and $\sigma\mathrm{Decl}_{\mathrm{sys}}$ to the position uncertainty at each epoch.  In the past, two main approaches have been employed \citep[e.g.,][]{chatterjee09a,deller12a,deller13a}:

\begin{enumerate}
\item Assume that $\sigma\mathrm{RA}_{\mathrm{sys}}$ and $\sigma\mathrm{Decl}_{\mathrm{sys}}$ are independent Gaussian-distributed variables, and estimate their variance by choosing values that give a reduced $\chi^2$ of 1.0 for the astrometric fit; or 
\item Perform a bootstrap fit \citep[e.g.][]{efron91a} to obtain realistic estimates for the uncertainties on the fitted parameters despite the fact that the uncertainties of the input positions are underestimated.  We conducted 100,000 trials, where each trial was conducted by sampling with replacement from the 8 position measurements (where each measurement consists of a right ascension and declination pair).
\end{enumerate}

In addition to applying these previous approaches, we also used a third technique for estimating $\sigma\mathrm{RA}_{\mathrm{sys}}$ and $\sigma\mathrm{Decl}_{\mathrm{sys}}$, which makes use of the apparent positional wander {\em within} a single observation.  In this procedure, for each epoch the variance in the right ascension and declination positions was calculated from the sub-divided, 30-minute time resolution datasets, and used to set $\sigma\mathrm{RA}_{\mathrm{sys}}^{2}$ and $\sigma\mathrm{Decl}_{\mathrm{sys}}^2$ respectively for the that epoch.  Accordingly, the estimated systematic error contribution varied from epoch to epoch.  If the apparent shifts on short timescales are a good proxy for the mean residual offset over a 4 hr period, then this approach would provide a more accurate estimation of the systematic errors than assuming them to have a Gaussian distribution with a fixed variance (the first appoach described above).

\begin{figure}
\begin{center}
\begin{tabular}{c}
\includegraphics[width=0.49\textwidth]{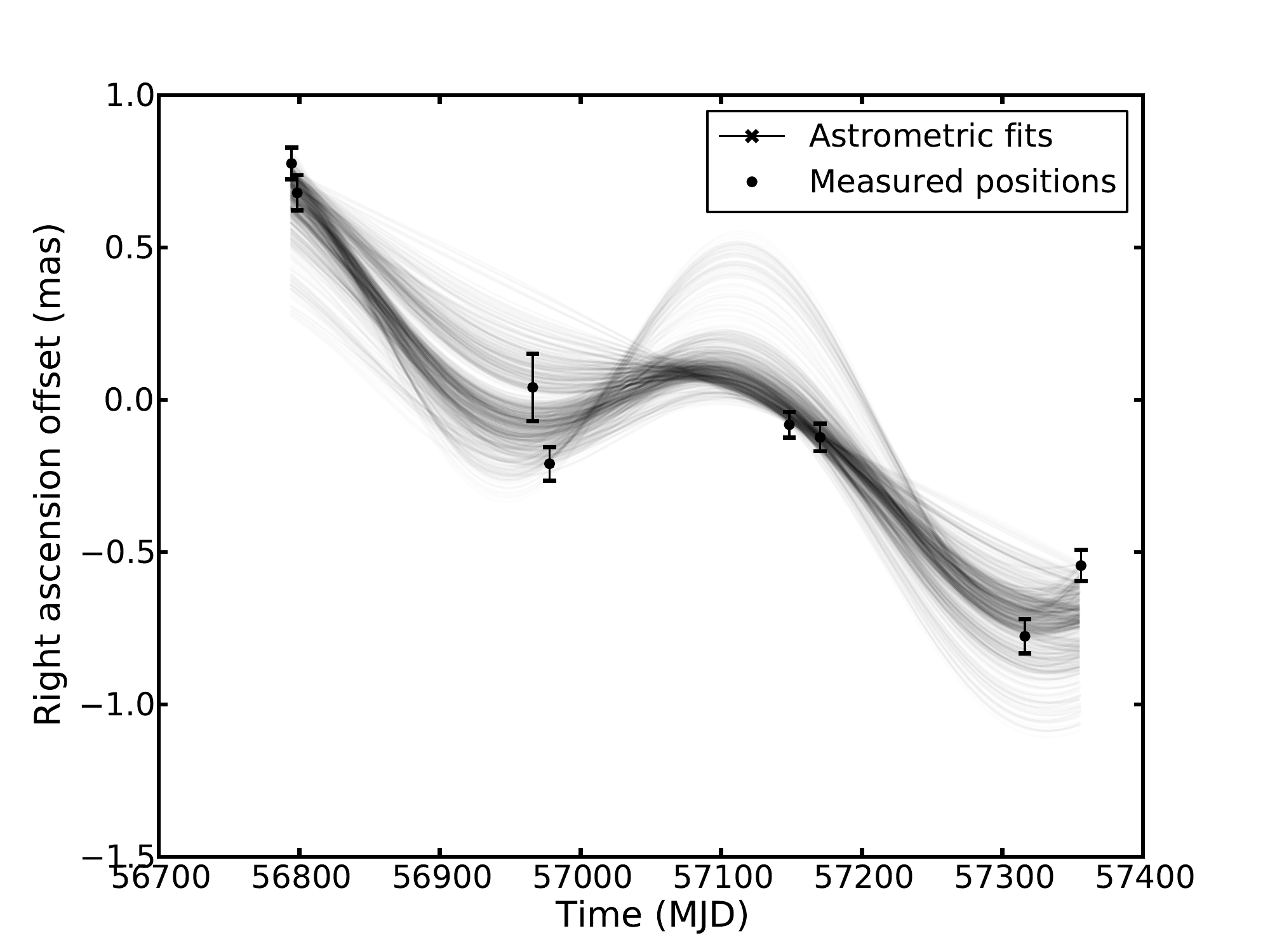} \\
\includegraphics[width=0.49\textwidth]{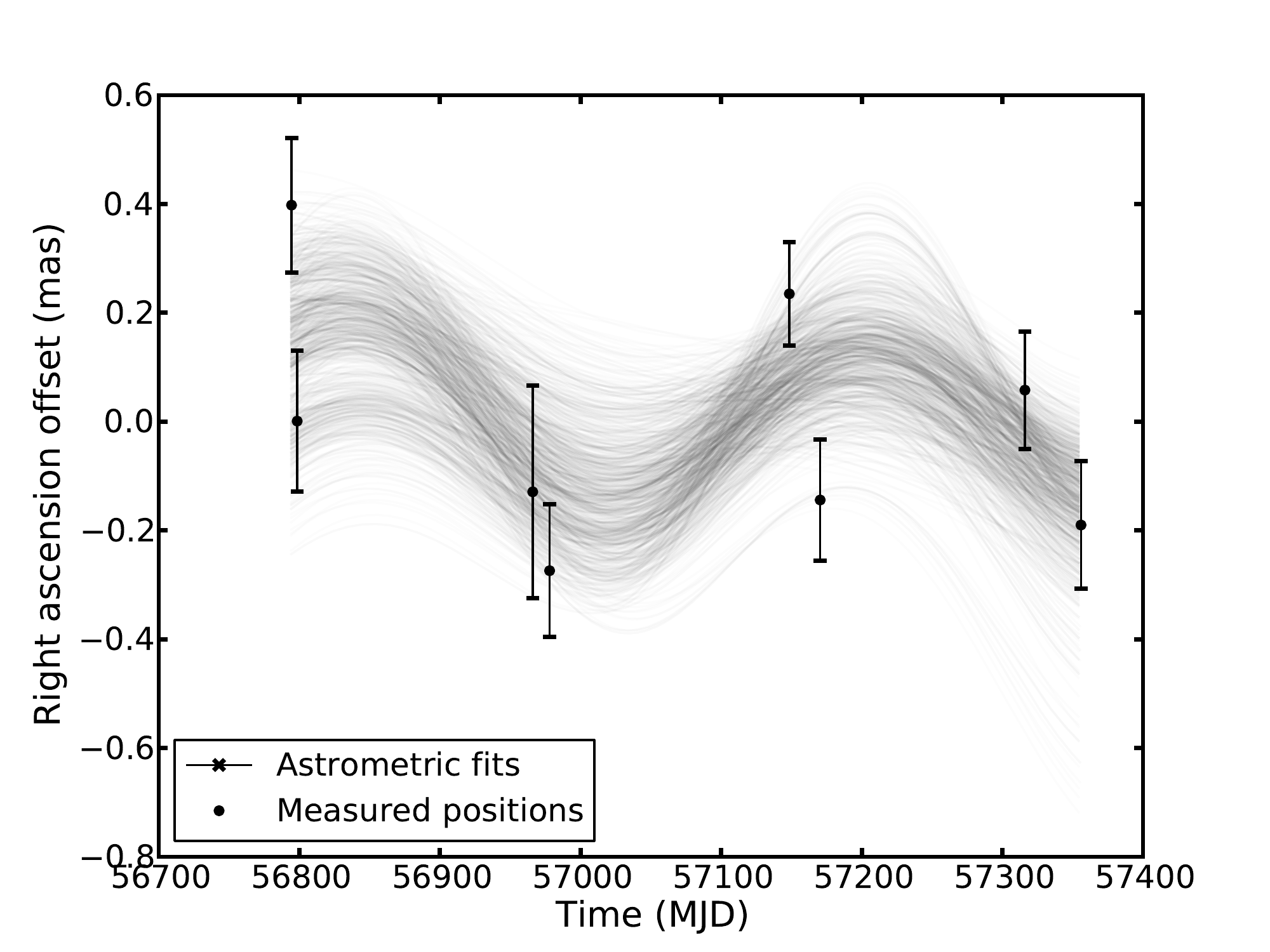} 
\end{tabular}
\end{center}
\caption{\label{fig:bootstrapfit}The bootstrap astrometric fit for PSR B1913+16, where the top panel and bottom panels show offset from the reference position (at MJD 57050) in right ascension and declination respectively. The resultant fit for of each of the 100,000 bootstrap trials is plotted in light grey.}
\end{figure}

We compared the results of all three approaches and found that the fitted astrometric values (reference position, proper motion, and parallax) all agreed to $\lesssim0.5\sigma$, while the estimated uncertainty values varied by up to $\sim$30\%.  
We report the results of the bootstrap fit, which gave the most conservative error bars for the fitted parameters of the three approaches.  The results are reported in Table~\ref{tab:fit} and plotted in Figure~\ref{fig:bootstrapfit}.  Absolute positions are provided in the International Celestial Reference Frame (ICRF) and are determined using the out of beam phase reference source 1911+1611, whose position was taken as 19$^\mathrm{h}$11$^\mathrm{m}$58$^\mathrm{s}$.257403$+$16$^\mathrm{d}$11'46''.86517.

\begin{deluxetable}{lr}
\tabletypesize{\small}
\tablewidth{0.47\textwidth}
\tablecaption{\label{tab:fit}Fitted astrometric parameters and 68\% 
confidence intervals\tablenotemark{a} for PSR B1913+16.}
\startdata
Right Ascension (J2000)\tablenotemark{b}				& 19:15:27.9986 $\pm$ 0.0001 \\
Declination (J2000)\tablenotemark{b}				& 16:06:27.381   $\pm$ 0.002 \\
Right Ascension offset (mas)\tablenotemark{c}  		& $-$775182.05 $\pm$ 0.04 \\
Declination offset (mas)\tablenotemark{c}				& $-$746452.86 $\pm$ 0.06 \\
Position epoch (MJD)							& 57050 \\
R.A. proper motion $\mu_{\alpha}$	(mas yr$^{-1}$)		& $-$0.77$^{+0.16}_{-0.06}$ \\
Decl. proper motion $\mu_{\delta}$	(mas yr$^{-1}$)		&      0.01$^{+0.10}_{-0.17}$ \\
Parallax $\pi$ (mas)	 				                         &       0.24$^{+0.06}_{-0.08}$ \\
\enddata
\tablenotetext{a}{Quoted uncertainties include systematic error estimates, and
are derived from a bootstrap fit (see text).}
\tablenotetext{b}{Uncertainty in absolute position is dominated by the absolute position uncertainty for the in-beam calibrator.}
\tablenotetext{c}{The relative offset between B1913+16 and the in-beam calibrator reference position (which is much more precise than the absolute position).}
\end{deluxetable}

\subsection{Estimating the distance to PSR B1913+16 from the VLBI parallax}
\label{sec:distance}
\begin{figure}
\begin{center}
\begin{tabular}{c}
\includegraphics[width=0.49\textwidth,trim=0 0.5cm 0  1.0cm, clip]{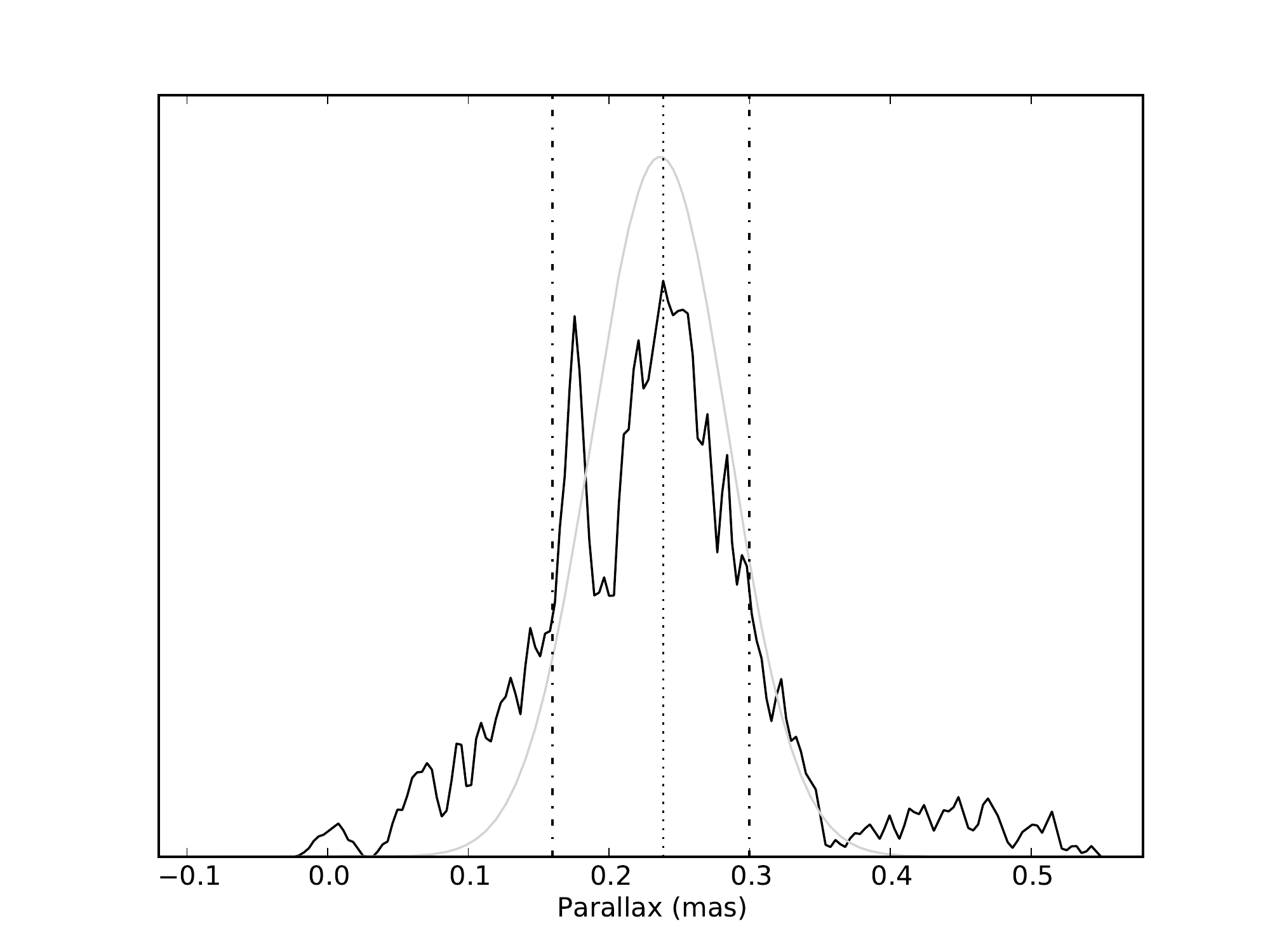}
\end{tabular}
\end{center}
\caption{\label{fig:parallaxhistogram}The histogram of parallax values resulting from the bootstrap fit (black solid line).  
The best-fit parallax  value is indicated with a vertical dashed line, and the 68\% confidence interval is indicated with vertical dash-dot lines. 
The distribution is not well approximated as a Gaussian, exhibiting a secondary peak; the spiky structure in the distribution is an unavoidable 
result of bootstrap fitting with small number of input samples possessing non-Gaussian errors.
The light grey line shows the Gaussian probability confidence interval resulting 
from a least-squares fit to the astrometric positions after including an estimate for $\sigma\mathrm{RA}_{\mathrm{sys}}$ and 
$\sigma\mathrm{Decl}_{\mathrm{sys}}$ that results in a reduced $\chi^2$ of 1.0; if this method were used, the parallax uncertainty would likely
be underestimated.}
\end{figure}

In order to improve the PSR B1913+16 test of relativistic gravitation, the most important quantity we wish to extract 
from our astrometric fit is the pulsar distance, which is constrained by the parallax.  A simple inversion of the 
measured parallax and its upper and lower limits, gives a nominal 68\% distance confidence interval of $d = 4.2^{+2.0}_{-0.9}$ kpc.  However, 
this result needs to be interpreted with caution for two reasons:

\begin{enumerate}
\item The distribution of parallax values obtained from the bootstrap fit is somewhat non-gaussian, as can be seen in 
Figure~\ref{fig:parallaxhistogram}; and
\item The significance of the parallax measurement is relatively low ($\sim$3$\sigma$), meaning that biases can 
substantially affect the distance inferred from the measured parallax \citep[e.g.,][]{verbiest10a,verbiest12a,igoshev16a}.
\end{enumerate}

\begin{figure}
\begin{center}
\begin{tabular}{c}
\includegraphics[width=0.49\textwidth,trim=0 1.0cm 0  0, clip]{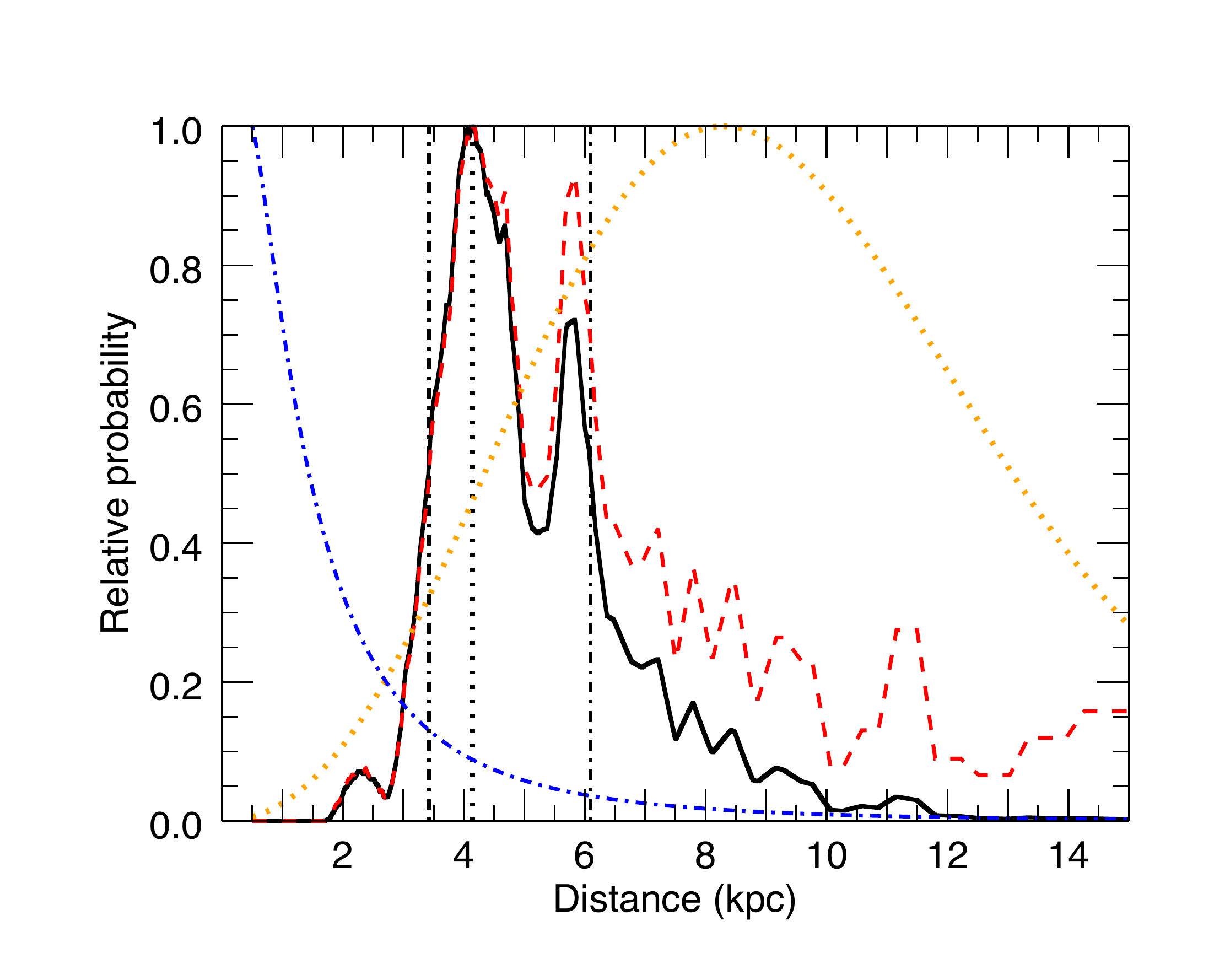}
\end{tabular}
\end{center}
\caption{\label{fig:disthistogram}The probability distribution of the pulsar distance (black solid line), which is the product of the parallax probability distribution 
obtained from our bootstrap fit (red dashed line), the 
volumetric term (yellow dotted line), and the luminosity term (blue dashed-dot line),  as given by  \citet{verbiest10a,verbiest12a} and \citet{igoshev16a}.  For display purposes, all probability distributions are shown normalised to a peak relative probability of 1.0.  The most probable distance is shown by a vertical black dotted line, and the 68\% confidence interval for distance is shown by vertical black dash-dot lines.  A smoothing kernel of width 0.3 kpc was
applied to the parallax probability distribution to mitigate the large fluctuations on short distance scales that result from the 
small sample size available for the bootstrap.}
\end{figure}

The accuracy of a distance (and distance error) inferred from a single low-significance parallax is strongly influenced by the choice of priors 
used \citep[e.g.][]{bailer-jones15a}.  For radio pulsar parallaxes, the most important priors are the galactic pulsar population distribution and the 
radio pulsar luminosity function, both of which are relatively poorly constrained.  The resultant distance, therefore, is strongly influenced by the 
assumed (yet poorly known) pulsar population characteristics.  Somewhat fortunately, the two priors act in opposite directions: more pulsars are 
expected along the line of sight to PSR B1913+16 at $d>4.2$ kpc than at $d<4.2$ kpc, but a $\sim$0.45 mJy radio pulsar such as PSR 
B1913+16 is more likely to be located at $d<4.2$ kpc, than $d>4.2$ kpc. (See Fig. \ref{fig:disthistogram}.)  

For the purpose of exploring the impact on tests of general relativity, we derive the pulsar distance from
 the VLBI parallax, corrected  for the Lutz-Kelker bias 
 (\citealp{verbiest10a,verbiest12a}; as modified by \citealp{igoshev16a}),
 using our own parallax distribution function 
 (Figure~\ref{fig:parallaxhistogram}),    and the same galactic pulsar population distribution and radio pulsar luminosity function used by 
  \citet{verbiest10a,verbiest12a}.  This procedure modestly shifts our distance estimate to 
 $d=4.1^{+2.0}_{-0.7}$ kpc.  With this distance, our measured proper motion yields a transverse velocity 
  of $v_{\mathrm T} = 15^{+8}_{-4}$ km s$^{-1}$ in the solar barycentric reference frame.	

\section{Discussion}
\label{sec:discussion}

\subsection{Non-GR contributions to $\dot{P}_{\rm \,b}$}

\citet{damour91a} derived the ``galactic'' correction caused by kinematic effects, $\dot{P}_{\rm \,b}^{\rm gal}$,
 to the observed orbital period derivative,  $\dot{P}_{\rm b}^{\rm \,obs}$, which is required in order to 
 determine the  intrinsic orbital period
 derivative at the pulsar system's  barycenter, $\dot{P}_{\rm b}^{\rm \,intr}$:
\begin{equation}
\dot{P}_{\rm b}^{\rm \,intr}=\dot{P}_{\rm b}^{\rm \,obs} - \dot{P}_{\rm b}^{\rm \,gal};
\label{eqn:obstointr}
\end{equation}
where 

\begin{equation}
\dot{P}_{\rm b}^{\rm \,gal} = (P_{\rm b}/c) 
\left[ 
\hat{\textbf n}_{p\ \leftarrow\odot}\cdot\left(\vec{\textbf a}_p-\vec{\textbf a}_\odot\right) + \mu^2d
\right],
\end{equation}
\citep{nice95a}.
Here $\hat{\textbf n}_{p\ \leftarrow\odot}$ is the unit vector pointing from the solar system to the
pulsar, $\vec{\textbf a}_p$ and $\vec{\textbf a}_\odot$ are the centripetal acceleration
of the   pulsar and  solar   system  barycenters in the
galactic potential, and  $\mu$ and $d$ are the pulsar's proper motion and distance.  
For PSR B1913+16, with galactic latitude $b= 2\fdg1$, it is sufficient to consider
only the galactic planar components of  $\hat{\textbf n}_{p\ \leftarrow\odot}$.  Consequently,
\begin{align}
&\dot{P}_{\rm b}^{\rm \,gal} =\frac{P_{\rm b}}{c} \times \nonumber \\
&\left\{  -\frac{  \Theta_0^2}{R_0}   \left[  \cos l + \left( \frac{ \Theta_{\rm psr}  }{ \Theta_0 } \right)^2    
 {\frac{ (d/R_0) -\cos l}  {1 - 2  (d/R_0) \cos l + (d/R_0)^2} } \right]  \right\}  \nonumber \\
&+  \frac{P_{\rm b}}{c} \mu^2d, 
\label{eqn:pbdotgal}
\end{align}
\citep{damour91a};
where $ \Theta_0$ is the (circular) velocity of the Local Standard of Rest and $ \Theta_{\rm psr}$ is
the equivalent quantity at the pulsar; $R_0$ is the galactocentric radius of the solar system; $l$ is the
pulsar's galactic longitude; and $\mu$ is its proper motion. (Note that the magnitude of the
 galactocentric centripetal acceleration  at any point, $|a|= \Theta^2/R$, where $\Theta$ and $R$
 are the systemic galactocentric circular speed and radius at that point.) 
 
 \subsection{Using the VLBI measurements and galactic parameters to evaluate $\dot{P}_{\rm b}^{\rm \,intr}$}
 
 In order to evaluate Eq. \ref{eqn:pbdotgal} so as to  subtract off the influence of the ``galactic'' term from  
 $\dot{P}_{\rm b}^{\rm obs}$, one needs to know the values of the pulsar's distance and proper 
 motion, and of various galactic constants.  Our   pulsar VLBI measurements detailed in 
 \S\ref{sec:astrometricresults} and \S\ref{sec:distance}  supply the former quantities, while a number of other 
 experiments have determined the latter parameters, particularly $R_0$ and $ \Theta_0$, via a variety of 
 techniques.   A concerted VLBI parallax campaign targeting galactic star-forming regions \citep{reid14a} 
  considerably improved the precision of the galactic quantities:  $ \Theta_0 = 240 \pm 8$ km  s$^{-1}$, 
  $R_0 = 8.34 \pm 0.16$ kpc, and  $d \Theta / d R =  -0.2 \pm 0.4$  km  s$^{-1}$ kpc$^{-1}$.  Two more recent 
  works have critically  and comprehensively 
  reviewed the galactic parameter determinations. \citet{bland-hawthorn16a} conclude that $R_0 = 8.2 \pm 0.1$ kpc and 
 $ \Theta_0 = 238 \pm 15$ km s$^{-1}$; while \citet{degrijs16a}, focusing principally on $R_0$,
 found its value to be $8.3\ (\pm 0.2$ [statistical] $\pm 0.4$ [systematic]) kpc,  with an implied $ \Theta_0$ from
 recent measurements of 270 $\pm$ 14 km s$^{-1}$.  Newer ``visual'' binary measurements of S star orbits at the 
 galactic center by \citet{boehle16a} suggested a significantly lower value of $R_0$, but when
 these measurements were incorporated into similar ones by \citet{gillessen17a}, the discrepancy 
 disappeared.  
 
 We calculated the value of $\dot{P}_{\rm b}^{\rm \,gal}$ based upon each of the above works, with an assumption
 of a flat rotation curve in each case except that of \citet{reid14a}, whose rotation curve slope was explicitly given (and
 still consistent with zero to within its errors as listed above.)  Most of them yield a similar value of 
 $\dot{P}_{\rm b}^{\rm \,gal}$, so our choice of a single best rotation curve is somewhat arbitrary.  Nevertheless,
these galactic parameters are  covariant.  Therefore,
we adopted the \citet{reid14a} galactic parameters; due  to their explicit simultaneous evaluation of all {\it{three}} relevant
parameters $R_0,   \Theta_0, and ~d \Theta / d R$; the latter of which affects the value of $ (\Theta_{\rm PSR}/ \Theta_0)$ 
in Eq. \ref{eqn:pbdotgal}.
 
These quantities, along with our newly determined pulsar distance  and proper motion  (summarized 
in Table \ref{tab:derived}) and its known galactic longitude
 $l=50\fdg0$,  yield $\dot{P}_{\rm b}^{\rm \,gal}= -(0.008^{+0.009}_{-0.003} ) \times 10^{-12}\ {\rm s/s}$.
  Inserting
 this value and  $\dot{P}_{\rm b}^{\rm \,obs}=-(2.423 \pm 0.001) \times 10^{-12}$ from \citet{weisberg16a} into
 Eq. \ref{eqn:obstointr}, we find that $\dot{P}_{\rm b}^{\rm \,intr}=  -(2.415  ^{+0.003}_{-0.009} ) \times 10^{-12}$.
 If we normalize this corrected, ``intrinsic'' quantity by the expected, GR-calculated value of 
 $\dot{P}_{\rm b}^{\rm \,GR} =  \left(-2.40263 \pm 0.00005 \right) \times 10^{-12}$ \citep{weisberg16a}, we find
 \begin{equation}
 \frac{ \dot{P}_{\rm b}^{\rm \,intr}  }{ \dot{P}_{\rm b}^{\rm \,GR}  }= 1.005 ^{+0.001}_{-0.003} .
  \end{equation}
    
In contrast, \citet{weisberg16a} determined a value of $0.9983 \pm 0.0016$.  They calculated the galactic 
correction of Eq.  \ref{eqn:obstointr}
 using  a pulsar distance estimate from \citet{weisberg08a}, as described in the next section.

\subsection{Comparison of  PSR B1913+16  distance estimates}

At the time of the \citet{weisberg16a}  $\dot{P}_b^{\rm gal}$ analysis, the best estimate of the pulsar distance 
was given by \citet{weisberg08a}. In the absence of
a {\it{direct}} distance measurement, these authors used measurements of the distance and mean electron
density toward pulsars whose lines of sight were near to B1913+16's, to determine its distance
based on its dispersion measure.  This procedure is akin to that undertaken with
global galactic electron density models \citep[e.g.,][which predicts a distance of 5.9~kpc]{cordes02a}, except that it used newer data and
was performed over a more limited region.  \citet{weisberg08a} found $d= 10.0 \pm 3.2\ (R_0/8.5\ {\rm kpc})$ kpc. 
Taking $R_0 = 8.34 \pm 0.16$ kpc \citep{reid14a} gives $d= 9.8 \pm 3.1\ $kpc, which was the estimate
used for the galactic correction $\dot{P}_{\rm b}^{\rm \,gal}$ in \citet{weisberg16a}.  Meanwhile, the newer ``YMW16" 
galactic electron density model \citep{yao17a} yields a DM distance of $d_{\mathrm{YMW}}=5.25$~kpc and estimates that the predicted distance
$d_{\mathrm{YMW}}$ will fall in the range 
$0.55\times d_{\mathrm{actual}} < d_{\mathrm{YMW}} < 1.45\times d_{\mathrm{actual}}$ in 68\% of cases.

While more precise than the \citet{weisberg08a} and \citet{weisberg16a} distance estimates,  the VLBI distance presented here
is in mild tension with the value
of $d=7.2$ kpc that is obtained by imposing the constraint $\dot{P}_{\rm b}^{\rm \,obs} = \dot{P}_{\rm b}^{\rm \,gal} + \dot{P}_{\rm b}^{\rm \,GR}$
and solving for $d$.  The most probable VLBI distance differs from this value of 7.2~kpc by 1.6\,$\sigma$.  However, as shown in Figure~\ref{fig:disthistogram},
the distance probability distribution is quite asymmetric, with the VLBI distance more likely to be underestimated than over-estimated.

\begin{deluxetable}{lrcc}
\tablecolumns{4}
\tablecaption {\label{tab:derived}Comparison of PSR B1913+16 VLBI astrometry to previous results.}
\tablehead{
\colhead{ Parameter}  				&   \colhead{VLBI} 			& \colhead{Weisberg \&} 	& \colhead{Yao et} 		\\
								&   \colhead{This work} 		& \colhead{Huang (2016)}	& \colhead{al. (2017)} 	\\}
\startdata
\cutinhead{Fitted Parameters} 
$\mu_{\alpha}$ (mas  yr$^{-1}$)  \dotfill	& $-$0.77$^{+0.16}_{-0.06}$ 	& $-1.23 \pm 0.04$ 				& 	\\ 
$\mu_{\delta} $ (mas  yr$^{-1}$)  \dotfill 	&       0.01$^{+0.10}_{-0.17}$ 	& $-0.83 \pm 0.04$ 	 			&	\\  
$\pi$   (mas)                                  \dotfill  	& 0.24$^{+0.06}_{-0.08}$  	& --- 			          			&	\\
\cutinhead{Derived Parameters} 
$d$  (kpc)   \dots 			                 & $4.1^{+2.0}_{-0.7}$  		& $9.8 \pm 3.1$\tablenotemark{a}   	& 5.25\tablenotemark{b} \\
$v_{\mathrm T}$ (km s$^{-1}$) \dots  	& $15^{+8}_{-4}$ 			& $69^{+25}_{-24}$ 	 & \\
\enddata 
\tablenotetext{a}{Analysis from \citet{weisberg08a}}
\tablenotetext{b}{The relative distance uncertainty is 0.9 (95\% C.I.).}
\end{deluxetable}

An improvement to the tests of general relativity with PSR B1913+16 would require improving our astrometric
accuracy by a factor of 3, reducing the parallax uncertainty to $\lesssim$20 $\mu$as.  Simulations suggest that 
if our systematic error budget remained unchanged, then a further 15-20 astrometric epochs would be required 
to approach this level of parallax.  However, new techniques such as
the use of multiple primary calibrators \citep[``MultiView;"][]{rioja17a} offer the potential
to better model and remove the ionospheric contamination, reducing these systematics.  Moreover, the bulk of our observations
took place close to solar maximum, when ionospheric activity (the major contributor to our systematic
error budget) is highest.  Accordingly, a continuation of VLBI observations with a comparable number of 
epochs to that presented here could reach a $\sim$10\% 
distance uncertainty for PSR B1913+16, especially if they took place over the 
next few years closer to solar minimum.

\subsection{Comparison of PSR B1913+16 proper motion measurements}

\begin{figure}
\begin{center}
\begin{tabular}{c}
\includegraphics[width=0.49\textwidth]{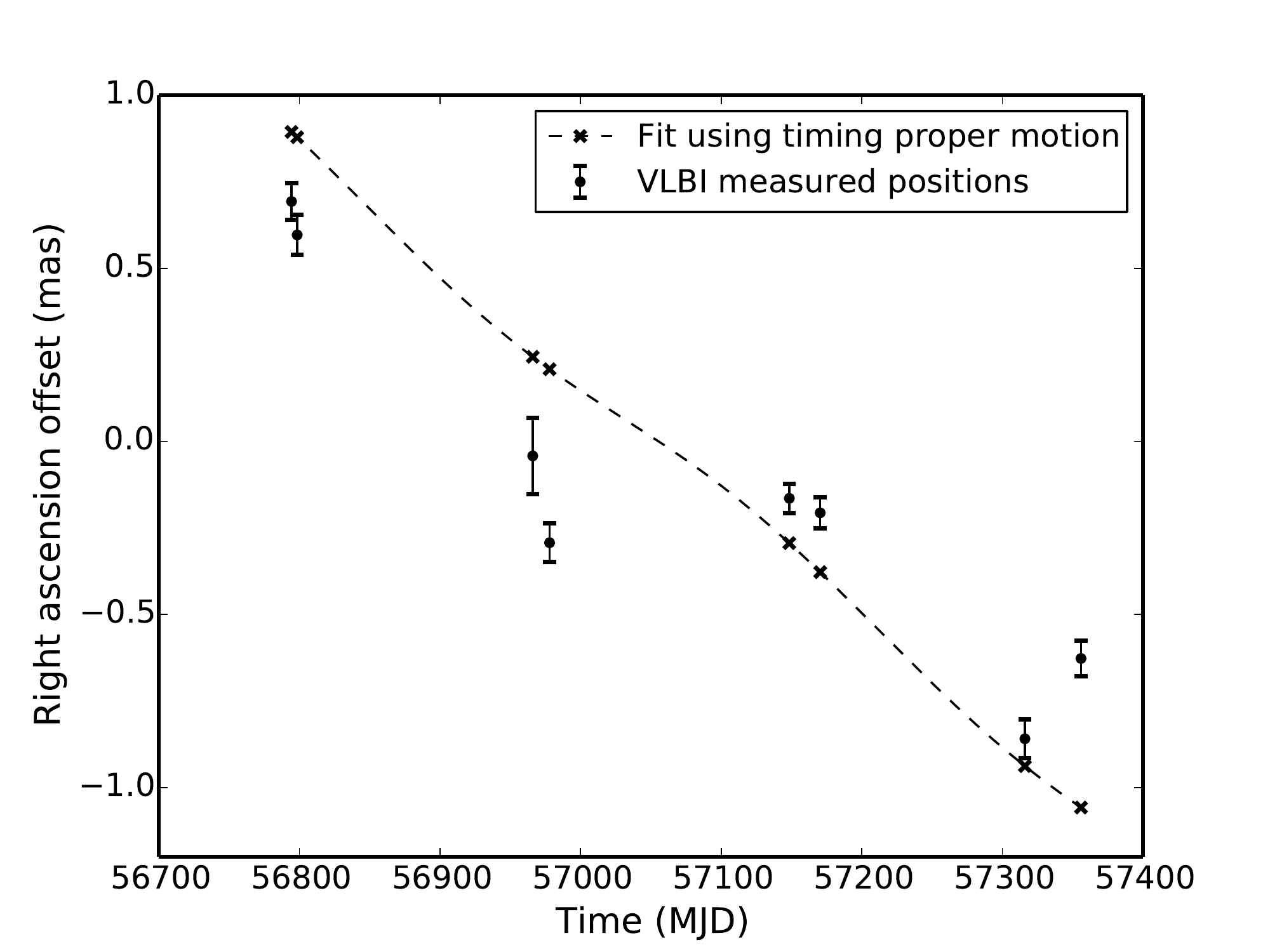} \\
\includegraphics[width=0.49\textwidth]{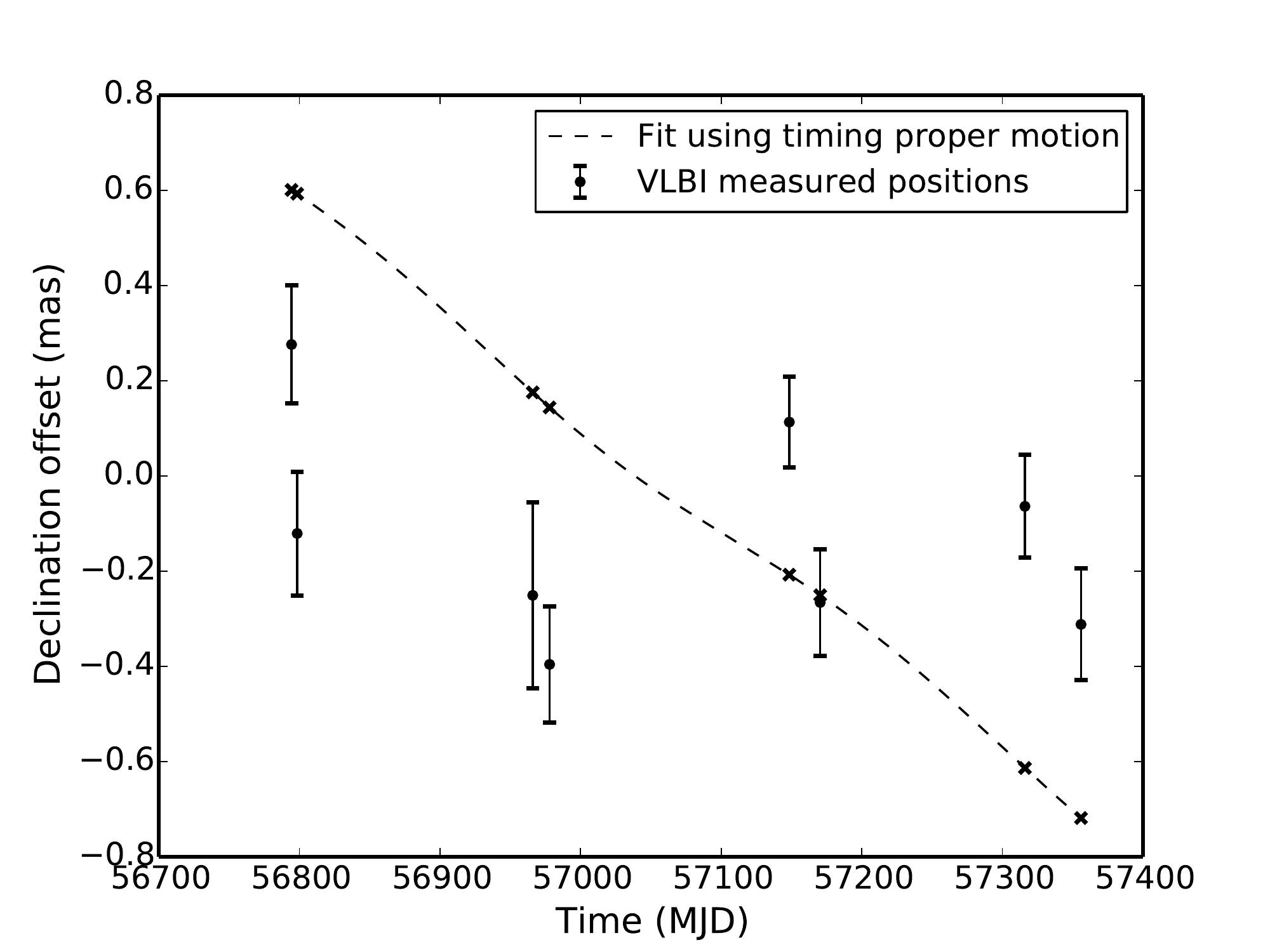} 
\end{tabular}
\end{center}
\caption{\label{fig:vlbitimingpm}The VLBI fit (for parallax and reference position only) after fixing the proper motion to the timing values from \citet{weisberg16a}.
Offset in right ascension is shown versus time in the top panel, and declination offset versus time in the bottom panel.  The fit is
much poorer than when proper motion is also fit, as shown in Figure~\ref{fig:simplefit} and Table~\ref{tab:fit}.}
\end{figure}

While the tests of general relativity presented above are still primarily limited by
the contribution of differential galactic rotation and hence by the pulsar distance 
uncertainty, it is also interesting to consider the difference in proper motion
obtained by VLBI astrometry versus pulsar timing (which contributes to the 
smaller Shklovskii term in $\dot{P}_{\rm b}^{\rm \,gal}$).  As shown in Table~\ref{tab:derived},
the VLBI proper motion results differ from the timing values presented in \citet{weisberg16a} by $>$4$\sigma$.
The discrepancy is highlighted in Figure~\ref{fig:vlbitimingpm}, which shows the best VLBI fit when the proper motion
is held fixed at the timing values. The reduced $\chi^2$ increases to 25.6 (from 4.7), and the best-fit parallax changes to 0.04 mas. 

Recently, \citet{deller16a}
showed that the uncertainties of proper motion estimates derived by timing for a number of millisecond pulsars were substantially
underestimated.  The pulsars with the largest errors in timing proper motion were preferentially located at low ecliptic latitudes,
where timing observations have reduced sensitivity to position changes in ecliptic latitude, and furthermore the effect of the solar wind
on dispersion measure variations is greatest.  PSR B1913+16 is however located at an ecliptic latitude of 38\degrees, where these effects
should be small or negligible.

Using a longer timing dataset, \cite{arzoumanian17a} found good agreement between VLBI and timing proper motions for two millisecond pulsars, including one of the discrepant cases identified by \citet{deller16a}.  They suggest that the previous disagreement was due to underestimation of the effect of bias due to timing noise in those data.  Here we explore the possibility of noise bias in the timing proper motion of PSR B1913+16.

The B1913+16 timing observations of \citet{weisberg16a} consist of two types of observing sets: (1) many short-term campaigns, in which all pulsar orbital phases were observed over a short period of time, typically separated by one or more years, and (2) two campaigns of observations spread over a year or more, one in 1985-1988, and one in 2004.  In these latter two campaigns, removal of the annual variation of time-of-flight delays across the Earth's orbit depends very sensitively on the position of the pulsar; in effect, then, they yield pulsar position measurements in 1985-1988, and again in 2004; the combination of these two sets of position measurements yields the proper motion.  (In practice, the timing proper motion was actually measured while simultaneous fitting for many other pulsar timing parameters; nevertheless, this division of the data set into campaigns is a useful paradigm for analyzing the proper motion measurement.)

A plausible explanation for the discrepancy between the VLBI and timing proper motion is that the timing in these around-the-year observations was influenced by noise in the observational time series on the $\sim$1~yr time scale needed to make timing position measurements.  We compared the original \citet{weisberg16a} timing solution with a timing solution in which the proper motion is fixed at the VLBI value, and we found differences in timing residuals with approximately annual time delays with amplitude $\sim 15~\mu$s over the course of these campaigns.  The resulting fit was marginally worse, with a reduced $\chi^2= 1.11$ for 9227 degrees of freedom, versus  reduced $\chi^2= 1.00$ for 9225 degrees of freedom, and most fitted parameters shifted by (1--2)\,$\sigma$ between the two fits.  Next we explore two possible sources that could introduce noise into the timing data at this level.

One possible source of such noise is pulsar rotation irregularities (commonly called ``timing noise'').  In fact, \citet{weisberg16a} observed significant timing noise in PSR B1913+16, which they modeled as ten higher-order spin frequency derivatives.  Their data set was approximately 32~yr long, so this effectively smoothed the data on time scales of roughly 32/10=3.2~yr and longer.  However, timing noise on a $\sim$1~yr time scale, if present, would have remained in the data.  Thus, this cannot be ruled out as a source of noise in the B1913+16 data set and the biased proper motion measurement.

A second possible source of such noise is variations in the dispersion measure (DM) of the pulsar due to the motion of the Earth-pulsar line-of-sight through the interstellar medium.  DM variations are a common feature of high precision pulsar data sets.  Mimicking the $\sim 15~\mu$s timing residual at a typical observing frequency of 1400\,MHz would require DM variations of $\Delta{\rm DM} = 0.007~{\rm pc\ cm}^{-3}$ on a time scales of 1~yr.  Prior to 2003, the timing of B1913+16 was done at a single observing frequency, so timing variations due to DM variations cannot be distinguished from timing noise intrinsic to the pulsar rotation.  After 2003, observations were made over a range of radio frequencies from 1170 to 1570~MHz, potentially allowing epoch-by-epoch measurement of DM.  We searched for DM variations in the 2004 data, but the results were inconclusive at the level of interest, largely because the measurement uncertainty at any given epoch was around $0.003~{\rm pc\ cm}^{-3}$, not much smaller than the DM variations needed to bias the observed proper motion.

To further consider the plausibility of DM variations of order $0.007~{\rm pc\ cm}^{-3}$, we sought out pulsar data sets for pulsars with DM similar to PSR~B1913+16.  A relatively close match is PSR~J1747-4036: it has a DM of $152.96~{\rm pc\ cm}^{-3}$, comparable to the DM of $168.35~{\rm pc\ cm}^{-3}$ for PSR~B1913+16, and a proper motion of $\sim 2$~mas/yr, only about twice that of PSR~B1913+16.  A measured DM time series for PSR~J1747$-$4036 is given in Figure 28 of \cite{arzoumanian17a}; it shows variations of around $0.006~{\rm pc\ cm}^{-3}$ on time scales of several years.  This is only moderately smaller than that needed to explain the PSR~B1913+16 proper motion bias, and it adds credibility to the thesis that DM variations may play at least a partial role in explaining the PSR~B1913+16 proper motion timing bias.

However, the VLBI proper motion is also subject to potential systematic influences that could lead to an under-estimated error.
The chief concern is evolution in the structure of the calibrator source J191621.8+161853.  Radio AGN can possess jets which
evolve on the timescales of years, leading to a shift in the source centroid position.  Since we by necessity use a static model for 
the calibrator source structure, any changes will manifest as a shift in the relative position of our target, PSR B1913+16.  However, 
very few radio AGN exhibit apparent centroid motion at a level greater than our proper motion uncertainty.  \citet{moor11a} studied
a sample of objects comparable to our calibrator source and found that 80\% 
exhibit apparent motion of less than 20 \uas\ yr$^{-1}$.  Just one source
out of 61 had an apparent motion $>0.1$ mas yr$^{-1}$, the level necessary to match or exceed our nominal VLBI proper motion uncertainty.

 A continued VLBI campaign over several years would reduce the VLBI
proper motion uncertainty to below that of the timing value, which will aid in pinpointing the cause of the discrepancy.

\subsection{The space velocity of PSR B1913+16}

The total proper motion of the pulsar, $\mu=0.77$~mas~yr$^{-1}$, 
combined with its 
distance of $d=4.1^{+2.0}_{-0.7}$~kpc, implies a transverse
speed of $v=15^{+8}_{-4}$~km~s$^{-1}$ in the Solar System Barycentric (SSB) frame.   This is a remarkably
low speed.  Because PSR~B1913+16 is several kiloparsecs
from the Sun, the motion of the standard of rest at the pulsar's position 
in the Galaxy differs significantly from that of the Local Standard of Rest. 
This difference alone could result in apparent
transverse motion.  Accounting for both this 
difference (using the galactic structure model of \cite{reid14a})
and for
the solar motion relative to the Sun's Local Standard of Rest
\citep{schonrich10a}, 
a star at rest relative to the
standard of rest at the position of the pulsar would have an observed SSB transverse speed
$v=110^{+70}_{-30}$~km~s$^{-1}$.  The observed
transverse speed of the pulsar is much smaller than this, which can only be
explained by fortuitous cancellation of the motion of its standard
of rest relative to the SSB and the motion of the pulsar relative to
its own standard of rest.   Thus the motion of the pulsar relative to
its standard of rest is likely of order 100~km~s$^{-1}$.  The exact
direction depends on its unknown line-of-sight velocity, but it is
likely dominated by motion radially away from the galactic center.
\cite{weisberg10a} reached a similar conclusion; they used 
different proper motion and distance estimates, but they also found the
observed pulsar velocity to be smaller than expected from galactic rotation.

Because the pulsar is very close to the galactic plane (galactic
latitude $b=2.1^\circ$), it is possible to robustly measure
its velocity perpendicular to the galactic disk, as it is nearly
independent of the line-of-sight velocity and it is not biased
by galactic rotation.
After accounting for solar motion, we find the pulsar
velocity perpendicular to the galactic plane to be 
$v_z=22^{+7}_{-3}$~km~s$^{-1}$.  Because the pulsar was
discovered in a survey at low galactic latitudes
\citep{hulse75a}, this low value of $v_z$ may
be an observational selection effect.

The formation and evolution of double neutron star (DNS) binaries is
of interest for interpreting LIGO
sources such as GW170817 \citep{abbott17a}.
\cite{tauris17a} undertook a comprehensive Monte Carlo analysis
of this subject based on all available observational data on
 (DNS) binaries (pre-GW170817).
They found that
most DNS received small birth kicks, but a small number
of systems, including PSR~B1913+16, must have received larger kicks
\citep[see also, e.g.,][]{wong10a,beniamini16a}.
For B1913+16, they found that
the supernova explosion forming the second neutron star in the system must have
been accompanied by a particularly large kick. They calculated that the kick velocity was in the range
 185--465~km~s$^{-1}$ if 
the system is at a distance of 9.8~kpc \citep{weisberg10a}, or some 50--100~km~s$^{-1}$ smaller
if $d=$ 5.25~kpc \citep{yao17a}.  
Our own distance measurement  is consistent with \citet{yao17a} and thus favors the latter, still large, kick velocity estimate.
 A fuller analysis of the 3-dimensional motion of PSR~B1913+16 is beyond the scope of the
present paper.

\section{Conclusions}
\label{sec:conclusions}

We have conducted  the first astrometric measurements of PSR B1913+16 using VLBI, with the intention  of measuring its
parallax and proper motion. Our principal motivations were  to make robust determinations of these quantities
and to improve the precision of the galactic acceleration 
contribution to the observed orbital decay of binary pulsar B1913+16.   When estimating the pulsar distance, 
we have extended the approach taken in previous works to incorporate not only prior information on the
millisecond pulsar spatial and luminosity distributions, but also the asymmetric parallax probability distribution
obtained from our VLBI astrometric fit.

Our measured distance differs from the best-fit DM-derived value \citep{weisberg08a}  by $\sim$3$\sigma$,
while our measured proper motion differs from the previous best timing-derived value  \citep{weisberg16a} by
$>$4$\sigma$.  These new measurements provide a galactic acceleration-induced correction to the measured orbital
decay leading to a test of general relativistic gravitational radiation damping that is of similar precision and lower accuracy 
than previous results, with a formal discrepancy from general relativity of 1.6\,$\sigma$.  

Our measured distance and proper motion confirm earlier suggestions that the second neutron star formed in this binary system suffered 
an unusually large birth kick.

We plan to conduct additional timing
and interferometric measurements. In this fashion, we will  improve our
experimental precision and further quantify both our random and systematic errors, a process that should 
resolve the somewhat discrepant 
results of the two techniques and probably the implied discrepancy with
general relativity. We note that most other binary pulsar orbital decay measurements are consistent with general relativity
\citep{weisberg16a}.

\acknowledgements  The Long Baseline Observatory is a facility of the National Science Foundation operated under cooperative 
agreement by Associated Universities, Inc.  ATD is the recipient of an Australian Research Council Future Fellowship (FT150100415).
JMW acknowledges support from  U.S. National Science 
Foundation Grant  AST-1312843. Pulsar research at Lafayette College and Cornell University is supported by the NANOGrav Physics Frontiers Center, 
funded through National Science Foundation award number 1430284.
The authors thank Y. Huang, G. Janssen, and J. Verbiest for useful discussions.

\bibliographystyle{apj}

\begin{thebibliography}{}
\expandafter\ifx\csname natexlab\endcsname\relax\def\natexlab#1{#1}\fi

\bibitem[{{Abbott} {et~al.}(2017){Abbott}, {Abbott}, {Abbott}, {Acernese},
  {Ackley}, {Adams}, {Adams}, {Addesso}, {Adhikari}, {Adya}, {Affeldt},
  {Afrough}, {Agarwal}, {Agathos}, {Agatsuma}, {Aggarwal}, {Aguiar}, {Aiello},
  {Ain}, {Ajith}, {Allen}, {Allen}, {Allocca}, {Altin}, {Amato}, {Ananyeva},
  {Anderson}, {Anderson}, {Angelova}, {Antier}, {Appert}, {Arai}, {Araya},
  {Areeda}, {Arnaud}, {Arun}, {Ascenzi}, {Ashton}, {Ast}, {Aston}, {Astone},
  {Atallah}, {Aufmuth}, {Aulbert}, {AultONeal}, {Austin}, {Avila-Alvarez},
  {Babak}, {Bacon}, \& Others}]{abbott17a}
{Abbott}, B.~P., {Abbott}, R., {Abbott}, T.~D., {et~al.} 2017, \apj, 850, L40

\bibitem[Arzoumanian et al.(2017)]{arzoumanian17a} 
Arzoumanian, Z., Brazier, A., Burke-Spolaor, S., et al.\ 2017, arXiv:1801.01837 

\bibitem[{{Bailer-Jones}(2015)}]{bailer-jones15a}
{Bailer-Jones}, C.~A.~L. 2015, \pasp, 127, 994

\bibitem[{{Beniamini} \& {Piran}(2016)}]{beniamini16a}
{Beniamini}, P., \& {Piran}, T. 2016, \mnras, 456, 4089

\bibitem[Bland-Hawthorn \& Gerhard(2016)]{bland-hawthorn16a} 
Bland-Hawthorn, J., \& Gerhard, O.\ 2016, \araa, 54, 529 

\bibitem[Boehle et al.(2016)]{boehle16a} Boehle, A., Ghez, A.~M., 
Sch{\"o}del, R., et al.\ 2016, \apj, 830, 17 

\bibitem[{{Chatterjee} {et~al.}(2009){Chatterjee}, {Brisken}, {Vlemmings},
  {Goss}, {Lazio}, {Cordes}, {Thorsett}, {Fomalont}, {Lyne}, \&
  {Kramer}}]{chatterjee09a}
{Chatterjee}, S., {Brisken}, W.~F., {Vlemmings}, W.~H.~T., {et~al.} 2009, \apj,
  698, 250

\bibitem[{{Cordes} \& {Lazio}(2002)}]{cordes02a}
{Cordes}, J.~M., \& {Lazio}, T.~J.~W. 2002, ArXiv e-prints, 0207156,
  astro-ph/0207156

\bibitem[{{Damour} \& {Taylor}(1991)}]{damour91a}
{Damour}, T., \& {Taylor}, J.~H. 1991, \apj, 366, 501

\bibitem[de Grijs \& Bono(2016)]{degrijs16a} 
de Grijs, R., \& Bono, G.\ 2016, \apjs, 227, 5 

\bibitem[{{Deller} {et~al.}(2013){Deller}, {Boyles}, {Lorimer}, {Kaspi},
  {McLaughlin}, {Ransom}, {Stairs}, \& {Stovall}}]{deller13a}
{Deller}, A.~T., {Boyles}, J., {Lorimer}, D.~R., {et~al.} 2013, \apj, 770, 145

\bibitem[{{Deller} {et~al.}(2012){Deller}, {Camilo}, {Reynolds}, \&
  {Halpern}}]{deller12a}
{Deller}, A.~T., {Camilo}, F., {Reynolds}, J.~E., \& {Halpern}, J.~P. 2012,
  \apjl, 748, L1

\bibitem[{{Deller} {et~al.}(2009){Deller}, {Tingay}, {Bailes}, \&
  {Reynolds}}]{deller09b}
{Deller}, A.~T., {Tingay}, S.~J., {Bailes}, M., \& {Reynolds}, J.~E. 2009,
  \apj, 701, 1243

\bibitem[{{Deller} {et~al.}(2011{\natexlab{a}}){Deller}, {Brisken}, {Phillips},
  {Morgan}, {Alef}, {Cappallo}, {Middelberg}, {Romney}, {Rottmann}, {Tingay},
  \& {Wayth}}]{deller11a}
{Deller}, A.~T., {Brisken}, W.~F., {Phillips}, C.~J., {et~al.}
  2011{\natexlab{a}}, \pasp, 123, 275

\bibitem[{{Deller} {et~al.}(2011{\natexlab{b}}){Deller}, {Brisken},
  {Chatterjee}, {Cordes}, {Goss}, {Janssen}, {Kovalev}, {Lazio}, {Petrov}, \&
  {Stappers}}]{deller11b}
{Deller}, A.~T., {Brisken}, W.~F., {Chatterjee}, S., {et~al.}
  2011{\natexlab{b}}, in Proceedings of the 20th EVGA Meeting, held 29-31
  March, 2011 at Max-Planck-Institut f{\"u}r Radioastronomie, Bonn, Germany.
  Edited by Walter Alef, Simone Bernhart, and Axel Nothnagel, p.178, ed.
  W.~{Alef}, S.~{Bernhart}, \& A.~{Nothnagel}, 178


\bibitem[Deller et al.(2016)]{deller16a} Deller, A.~T., Vigeland, 
S.~J., Kaplan, D.~L., et al.\ 2016, \apj, 828, 8 

\bibitem[{{Efron} \& {Tibshirani}(1991)}]{efron91a}
{Efron}, B., \& {Tibshirani}, R. 1991, Science, 253, 390

\bibitem[Gillessen et al.(2017)]{gillessen17a} Gillessen, S., Plewa, P.~M., Eisenhauer, 
F., et al.\ 2017, \apj, 837, 30 

\bibitem[{{Greisen}(2003)}]{greisen03a}
{Greisen}, E.~W. 2003, Information Handling in Astronomy - Historical Vistas,
  285, 109

\bibitem[Hulse \& Taylor(1975)]{hulse75a} Hulse, R.~A., \& Taylor, J.~H.\ 1975, \apjl, 195, L51 

\bibitem[Igoshev et al.(2016)]{igoshev16a} Igoshev, A., Verbunt, F., \& Cator, E.\ 2016, 
\aap, 591, A123 

\bibitem[{{Kettenis} {et~al.}(2006){Kettenis}, {van Langevelde}, {Reynolds}, \&
  {Cotton}}]{kettenis06a}
{Kettenis}, M., {van Langevelde}, H.~J., {Reynolds}, C., \& {Cotton}, B. 2006,
  in Astronomical Society of the Pacific Conference Series, Vol. 351,
  Astronomical Data Analysis Software and Systems XV, ed. C.~{Gabriel},
  C.~{Arviset}, D.~{Ponz}, \& S.~{Enrique}, 497

\bibitem[{{Mo{\'o}r} {et~al.}(2011){Mo{\'o}r}, {Frey}, {Lambert}, {Titov}, \&
  {Bakos}}]{moor11a}
{Mo{\'o}r}, A., {Frey}, S., {Lambert}, S.~B., {Titov}, O.~A., \& {Bakos}, J.
  2011, \aj, 141, 178

\bibitem[{{Nice} \& {Taylor}(1995)}]{nice95a}
{Nice}, D.~J., \& {Taylor}, J.~H. 1995, \apj, 441, 429

\bibitem[{{Reid} {et~al.}(2014){Reid}, {Menten}, {Brunthaler}, {Zheng}, {Dame},
  {Xu}, {Wu}, {Zhang}, {Sanna}, {Sato}, {Hachisuka}, {Choi}, {Immer},
  {Moscadelli}, {Rygl}, \& {Bartkiewicz}}]{reid14a}
{Reid}, M.~J., {Menten}, K.~M., {Brunthaler}, A., {et~al.} 2014, \apj, 783, 130

\bibitem[{{Rioja} {et~al.}(2017){Rioja}, {Dodson}, {Orosz}, {Imai}, \&
  {Frey}}]{rioja17a}
{Rioja}, M.~J., {Dodson}, R., {Orosz}, G., {Imai}, H., \& {Frey}, S. 2017,
 \aj,153, 105

\bibitem[{{Sch{\"o}nrich} {et~al.}(2010){Sch{\"o}nrich}, {Binney}, \&
  {Dehnen}}]{schonrich10a}
{Sch{\"o}nrich}, R., {Binney}, J., \& {Dehnen}, W. 2010, \mnras, 403, 1829

\bibitem[{{Shepherd}(1997)}]{shepherd97a}
{Shepherd}, M.~C. 1997, in Astronomical Society of the Pacific Conference
  Series, Vol. 125, Astronomical Data Analysis Software and Systems VI, ed.
  G.~{Hunt} \& H.~{Payne}, 77

\bibitem[Shklovskii(1970)]{shklovskii70a} Shklovskii, I.~S.\ 1970,
 \sovast, 13, 562 

\bibitem[{{Tauris} {et~al.}(2017){Tauris}, {Kramer}, {Freire}, {Wex}, {Janka},
  {Langer}, {Podsiadlowski}, {Bozzo}, {Chaty}, {Kruckow}, {van den Heuvel},
  {Antoniadis}, {Breton}, \& {Champion}}]{tauris17a}
{Tauris}, T.~M., {Kramer}, M., {Freire}, P.~C.~C., {et~al.} 2017, \apj, 846,
  170

\bibitem[{{Taylor} \& {Weisberg}(1982)}]{taylor82a}
{Taylor}, J.~H., \& {Weisberg}, J.~M. 1982, \apj, 253, 908

\bibitem[{{Verbiest} {et~al.}(2010){Verbiest}, {Lorimer}, \&
  {McLaughlin}}]{verbiest10a}
{Verbiest}, J.~P.~W., {Lorimer}, D.~R., \& {McLaughlin}, M.~A. 2010, \mnras,
  405, 564

\bibitem[{{Verbiest} {et~al.}(2012){Verbiest}, {Weisberg}, {Chael}, {Lee}, \&
  {Lorimer}}]{verbiest12a}
{Verbiest}, J.~P.~W., {Weisberg}, J.~M., {Chael}, A.~A., {Lee}, K.~J., \&
  {Lorimer}, D.~R. 2012, \apj, 755, 39

\bibitem[Weisberg \& Huang(2016)]{weisberg16a} Weisberg, J.~M., \& Huang, 
Y.\ 2016, \apj, 829, 55 


\bibitem[Weisberg, Nice, \& Taylor(2010)]{weisberg10a} Weisberg, J.~M., Nice, D.~J., \& Taylor, J.~H.\ 2010, \apj, 722, 1030 

\bibitem[{{Weisberg} {et~al.}(2008){Weisberg}, {Stanimirovi{\'c}}, {Xilouris},
  {Hedden}, {de la Fuente}, {Anderson}, \& {Jenet}}]{weisberg08a}
{Weisberg}, J.~M., {Stanimirovi{\'c}}, S., {Xilouris}, K., {et~al.} 2008, \apj,
  674, 286

\bibitem[{{Wong} {et~al.}(2010){Wong}, {Willems}, \&
  {Kalogera}}]{wong10a}
{Wong}, T.-W., {Willems}, B., \& {Kalogera}, V. 2010, \apj, 721, 1689

\bibitem[Yao, Manchester, \& Wang(2017)]{yao17a} Yao, J.~M., Manchester, R.~N., \& Wang, N.\ 2017, \apj, 835, 29 

\end{thebibliography}

\end{document}